\documentclass[twocolumn,superscriptaddress]{revtex4}
\usepackage{graphicx}
\usepackage{epsfig}
\usepackage{amsmath}
\usepackage{amsfonts,amsbsy}
\usepackage{amssymb}
\usepackage{hyperref}
\usepackage{bm}
\usepackage{color}
\usepackage{enumerate}

\def\be{\begin{equation}}
\def\ee{\end{equation}}
\def\bea{\begin{eqnarray}}
\def\eea{\end{eqnarray}}

\def\fig#1{{Fig.~\ref{#1}}}

\newcommand{\bo}[1]{\boldsymbol{#1}}

\begin{document}
\title{Fine structures of azimuthal correlations of two gluons in the glasma}
\author{Hengying Zhang}
\affiliation{Key Laboratory of Quark and Lepton Physics (MOE) and Institute of Particle Physics,\\
Central China Normal University, Wuhan 430079, China}
\author{Donghai Zhang}
\affiliation{Key Laboratory of Quark and Lepton Physics (MOE) and Institute of Particle Physics,\\
Central China Normal University, Wuhan 430079, China}
\author{Yeyin Zhao}
\affiliation{Key Laboratory of Quark and Lepton Physics (MOE) and Institute of Particle Physics,\\
Central China Normal University, Wuhan 430079, China}
\author{Mingmei Xu}
\email{xumm@mail.ccnu.edu.cn}
\affiliation{Key Laboratory of Quark and Lepton Physics (MOE) and Institute of Particle Physics,\\
Central China Normal University, Wuhan 430079, China}
\author{Xue Pan}
\affiliation{School of Electronic Engineering, Chengdu Technological University, Chengdu 611730, China}
\author{Yuanfang Wu}
\email{wuyf@phy.ccnu.edu.cn}
\affiliation{Key Laboratory of Quark and Lepton Physics (MOE) and Institute of Particle Physics,\\
Central China Normal University, Wuhan 430079, China}
\date{\today}
\begin{abstract}
We investigate the azimuthal correlations of the glasma in p-p collisions at $\sqrt{s_{\mathrm{NN}}}=7$ TeV by using the CGC formalism. As expected, the azimuthal correlations show two peaks at $\Delta\phi=0$ and $\pi$ which represent collimation production in CGC. Beyond that, azimuthal correlations show fine structures, i.e. bumps or shoulders between the two peaks, when at least one gluon has small $x$. The structures are demonstrated to be associated with saturation momentum, and likely appear at transverse momentum around $2Q_{\mathrm{sp}}=1.8$ GeV$/$c.  
\end{abstract}

\maketitle
\section{Introduction}\label{sec:introduction}

Two-particle correlations in high energy collisions have been measured previously for a broad range of collision energies and colliding systems with the goal of understanding the underlying mechanism of particle production. Especially the study of two-particle azimuthal correlations provides important information for characterizing Quantum Chromo-dynamics (QCD), e.g. the mechanism of hadronization, possible collective effect and gluon saturation effect in these collisions.

One measurement of azimuthal correlations is performed using two-dimensional $\Delta\eta$-$\Delta\phi$ correlation functions. Here $\Delta\eta$ is the difference in pseudorapidity between the two particles and $\Delta\phi$ is the difference in their azimuthal angle (in radians). Another measurement of azimuthal correlations is projection of 2-D correlation functions onto $\Delta\phi$, i.e. $\Delta\phi$ distributions, usually at short-range rapidity (small $\Delta\eta$) and long-range rapidity (large $\Delta\eta$). 

In minimum bias p-p collisions, both in the data and in the Monte Carlo event generator PYTHIA, the two-dimensional  $\Delta\eta$-$\Delta\phi$ correlations in the intermediate $p_{\perp}$ range are dominated by two components~\cite{ridge-PHENIX,ridge-PHOBOS,CMS-pp-JHEP-2010}: a narrow peak at $(\Delta\eta,\Delta\phi)\approx(0,0)$ which can be understood as the contribution from jets; a ridge at $\Delta\phi\approx\pi$ extending over a broad range in $\Delta\eta$, interpreted as due to momentum conservation or away-side jets. If this azimuthal correlations are demonstrated as $\Delta\phi$ distribution at long-range rapidity ($|\Delta\eta|>2$), there is a peak at away side ($\Delta\phi\approx\pi$) and no peak at near side ($\Delta\phi\approx 0$) in minimum bias p-p collisions. However, in central heavy ion collisions, e.g. Au-Au~\cite{ridge-PHENIX,ridge-PHOBOS,ridge-STAR-1,ridge-STAR-2} and Pb-Pb~\cite{CMS-pPb-PbPb-PLB-2013} collisions, the azimuthal correlations measured in the intermediate $p_{\perp}$ range begin to show an elongated structure in $\Delta\eta$ direction at near side, known as the near-side ridge. The corresponding $\Delta\phi$ distribution at large $\Delta\eta$ starts to show a peak at near side which is absent in minimum bias p-p collisions. 

Later, high-multiplicity events in small systems, like p-p, p-Pb collisions at the LHC~\cite{CMS-pp-JHEP-2010,CMS-pPb-PLB-2013,CMS-pPb-PbPb-PLB-2013,ATLAS-pPb-PRL-2013,ALICE-pPb-PLB-2013,ALICE-pPb-PLB-2016} and d-Au, He-Au collisions at RHIC~\cite{ridge-RHIC-small-1,ridge-RHIC-small-2} also show near-side ridge phenomenon which is similar to that of the heavy-ion collisions. Careful subtraction of the azimuthal correlations in low-multiplicity events from those in high-multiplicity events, shows that the magnitude of the near-side peak is nearly identical to that of the away-side peak, known as double-ridge phenominon~\cite{ATLAS-pPb-PRL-2013,ALICE-pPb-PLB-2013,ALICE-pPb-PLB-2016}. Collimated production in azimuthal angle is a prominent feature at long-range rapidity. 

Azimuthal correlations at long-range rapidity in small systems can be explained by initial-state effects such as gluon saturations~\cite{cgc-NPA-2008,cgc-NPA-2010,cgc-PRD-I,cgc-PRD-II,cgc-PRD-III}, final-state parton-parton induced interactions~\cite{final-parton-interaction}, hydrodynamic flow~\cite{review-dusling,pp-hydro,mass-order-pp}, etc.  Among them, gluon saturation in the initial state is the most promising one because the dihadron azimuthal correlations at long-range rapidity calculated from a combination of glasma dynamics and BFKL (dominating the away side peak) agree well with data in p-p and p-Pb collisions over a very wide range of $p_{\perp}^{\mathrm{trig}}$, $p_{\perp}^{\mathrm{asc}}$ windows, centrality class and $\Delta\eta$ acceptance on a quantitative level~\cite{cgc-PRD-I,cgc-PRD-II,cgc-PRD-III}. It suggests that two-particle azimuthal correlations are sensitive to detailed dynamical features of CGC. 
 
Azimuthal, together with transverse momentum and rapidity, constitute the three-momentum of a particle. The distributions and correlations of azimuthal angle are necessarily influenced by transverse momentum and rapidity due to, e.g. momentum conservation. For the study of the rapidity and transverse momentum dependence of azimuthal correlations, CGC is unique as the following reasons.

\begin{figure}[t]
\begin{centering}
\par
\includegraphics[scale=0.74]{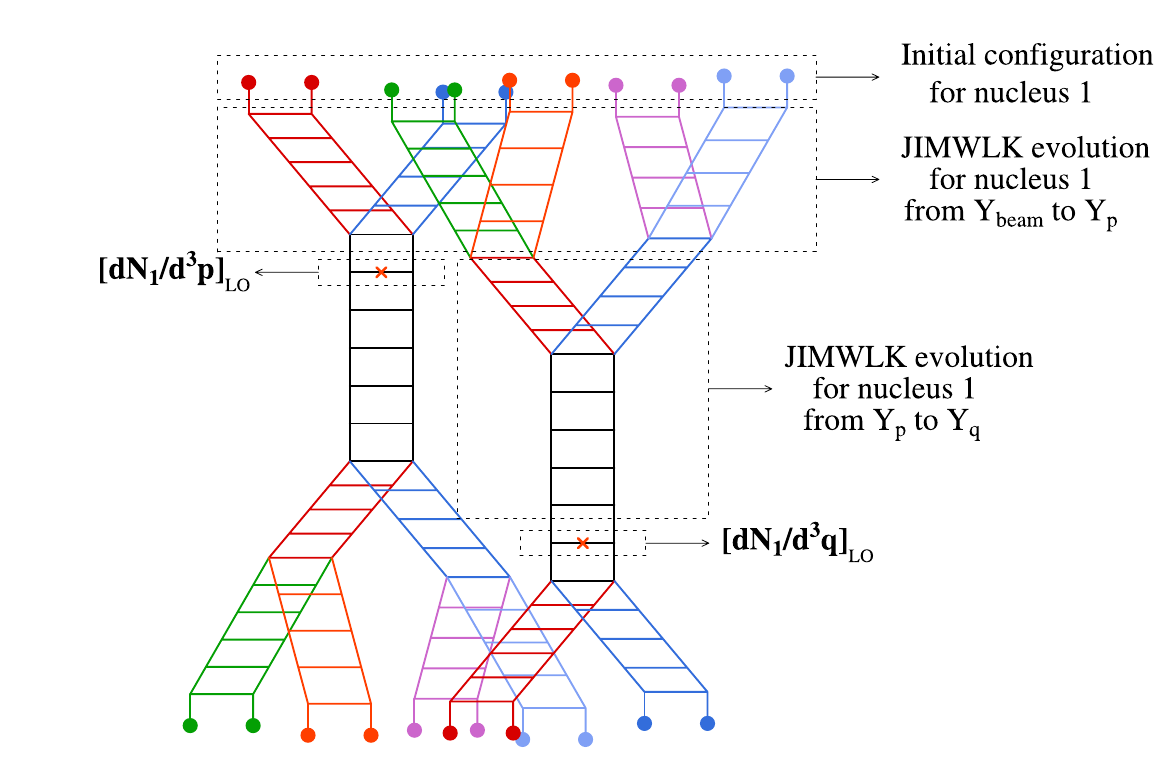}
\end{centering}
\caption{A schematic diagram of high energy QCD evolution.  Two crosses  mark two observed gluons with momentum $\textbf{p}$ and $\textbf{q}$. $Y_p$ and $Y_q$ denote their rapidities in the infinite momentum frame, i.e. $Y=\ln\frac{x_0}{x}$. The lower part of the figure, representing nucleus 2, is made up of identical evolution. Figure is from ref~\cite{cgc-NPA-2010}.}
\label{glasma-evolution}
\end{figure}

\begin{enumerate}[(1)]
\item{In CGC, gluons are progressively emitted from valence quarks by the ladder graph, shown in \fig{glasma-evolution}. Hence partons at different step have different Feynman $x$, i.e. the longitudinal momentum fraction in the projectile or target. According to the longitudinal momentum fraction, partons can be classified as fast partons with large $x$, whose longitudinal momentum $p^+$ is larger than a cutoff, and slow partons with small $x$, whose longitudinal momentum $p^+$ is smaller than the cutoff. Furthermore, Feynman $x$ is related to the rapidity. For a right-moving parton, its rapidity $y$ is related to its Feynman $x$ as $x=\frac{m_{\perp}}{\sqrt{s}}e^{y}$ in high energy limit. At $\sqrt{s}=7$ TeV and intermediate $p_{\perp}$, e.g. 2 GeV$/$c, the rapidity at $y=1$ corresponds to $x\sim 10^{-3}$, while $y=2$ corresponds to $x\sim 10^{-2}$. Gluons at the central rapidity region reflect the properties of the small-$x$ ($x< 10^{-3}$) degrees of freedom. Gluons with larger rapidity have larger $x$ in the right-moving projectile. When one parton has small rapidity and the other has forward large rapidity, the correlations between them reflect the correlations between slow gluons and fast gluons. Hence gluons at different rapidity location lie in different dynamical region, and the aizimuthal correlations between them may be different if they are chosen from different rapidity locations and reflect the dynamical regions they originate from.  }

\item{In CGC the transverse momentum distribution of saturated gluons (called uGD) has a peak, whose position is called the characteristic transverse momentum of gluons and equals to the saturation scale $Q_\mathrm{s}$. That means uGD only takes large value when $p_{\perp}$ is around  $Q_\mathrm{s}$. Since uGD determines the correlation function (as to be illustrated in section II), the azimuthal correlations should be $p_{\perp}$ dependent and sensitive. }

\end{enumerate}

 A systematic study of azimuthal anisotropy and its $p_{\perp}$ and $y$ dependence is important since it supplies special signals originating from CGC dynamics. It could provide crucial tests as to whether an initial-state interaction or a final-state interaction dominates the azimuthal correlation created in small systems. In this paper, we analyze the azimuthal correlations of p-p collisions at 7 TeV under CGC framework. Most of the existing results in references are limited by the range of rapidity gap, such as long-range rapidity ($|\Delta\eta|>2$) or short-range rapidity ($|\Delta\eta|<1$), no matter which rapidity location the gluons are situated in. However, the analysis in this paper is not limited by the range of rapidity gap, but by the rapidity location, i.e. small rapidity region (defined as $|y|\leqslant1$) and large rapidity region($|y|\geqslant 2$).

This paper is organized as follows. In section \ref{sec:review}, the definition of azimuthal correlation and some related formula of the single- and double-gluon inclusive production in CGC framework are given. Results of two-gluon azimuthal correlations are shown and discussed in section \ref{sec:results}, where the sensitivity to rapidity location and transverse momentum are carefully compared and some interesting correlation patterns are illustrated. Section
\ref{sec:summary} gives the summary.

\section{two-gluon azimuthal correlations from high energy QCD evolution}\label{sec:review}

In a high energy collision, both the projectile and the target are regarded as high parton density sources. When they pass through each other, strong
longitudinal color electric and color magnetic fields are formed.
The collection of the primordial fields at the early stage is called
the glasma. The framework to describe the physics of high parton densities and
gluon saturation is the Color Glass Condensate (CGC) effective field
theory
(EFT)~\cite{Gribov-1983,Iancu-2003,Weigert-2005,Gelis-2010}.
The effective degrees of freedom in this framework are color sources
$\rho$ at large $x$ and classical gauge fields $\mathcal{A}_\mu$ at
small $x$. The classical gauge field $\mathcal{A}_\mu$ is the
solution of Yang-Mills equations with a fixed configuration of color
sources. For a given initial configuration of color source, the distribution of sources and fields in the nuclear wavefunctions evolve with rapidity $Y$ or corresponding $x$, which is described by the JIMWLK renormalization group
equations~\cite{Jalilian-Marian-1997,Jalilian-Marian-PRD,Iancu-2001}, as shown in \fig{glasma-evolution}. Here $Y$ can be translated to the rapidity of a gluon in the center of mass frame $y$, i.e. $Y=\ln\frac{x_0}{x}=\ln x_0+\ln \frac{\sqrt{s}}{m_\perp}\pm y$. In \fig{glasma-evolution}, the dots at upper part represent the initial color sources for nucleus 1. The ladder graphs following it illustrate the high energy evolution of parton distribution, from $Y_{\mathrm{beam}}$ to $Y_p$ and from $Y_p$ to $Y_q$ by JIMWLK equation, with both radiation and scattering processes included. Two crosses  mark two observed gluons with momentum $\textbf{p}$ and $\textbf{q}$. At first glance the two gluons seem to be uncorrelated since they come from superficially disconnected diagrams. However, the two-gluon production should be calculated by averaging over the initial distribution of color sources, which introduces correlations.

Supposing two gluons are produced with transverse momentum
$\bo{p}_\perp$ and $\bo{q}_\perp$, and rapidity
$y_p$ and $y_q$, two-particle correlation is defined as

\begin{equation}
\label{eq:24}
\begin{split}
C(\bo{p}_\perp,y_p;\bo{q}_\perp, y_q)&= \frac{
\left\langle\frac{dN_2}{d^2\bo{p}_\perp dy_p d^2\bo{q}_\perp
dy_q}\right\rangle }{ \left\langle\frac{dN_1}{d^2\bo{p}_\perp
dy_p}\right\rangle \left\langle\frac{dN_1}{d^2\bo{q}_\perp
dy_q}\right\rangle }-1 \\  &=
\frac{
\left\langle\frac{dN^{\mathrm{corr.}}_2}{d^2\bo{p}_\perp dy_p d^2\bo{q}_\perp dy_q}\right\rangle
}{
\left\langle\frac{dN_1}{d^2\bo{p}_\perp dy_p}\right\rangle
\left\langle\frac{dN_1}{d^2\bo{q}_\perp dy_q}\right\rangle
},
\end{split}
\end{equation}
where $\left\langle\frac{dN_2}{d^2\bo{p}_\perp dy_p d^2\bo{q}_\perp
dy_q}\right\rangle$ and $\left\langle\frac{dN_1}{d^2\bo{p}_\perp
dy_p}\right\rangle$ are the double- and single-gluon inclusive
productions, and
$\left\langle\frac{dN^{\mathrm{corr.}}_2}{d^2\bo{p}_\perp dy_p
d^2\bo{q}_\perp dy_q}\right\rangle$ is the correlated double-gluon
production which subtracts the uncorrelated double-gluon production.
The leading log factorization formula reads~\cite{cgc-NPA-2008}
\begin{equation}\label{eq:llog}
\langle\mathcal{O}\rangle_{\mathrm{LLog}}=\int\left[D\rho_1\right]\left[D\rho_2\right]W[\rho_1]W[\rho_2]\mathcal{O}[\rho_1,\rho_2]_{\mathrm{LO}},
\end{equation}
where $\mathcal{O}[\rho_1,\rho_2]_{\mathrm{LO}}$ is the leading order
single or double particle inclusive distribution for a fixed
distribution of color sources, and the integration denotes an
average over different distribution of the color sources with the
weight functional $W[\rho_{1,2}]$. In general, $W[\rho_{1,2}]$
encodes all possible color charge configurations of projectile and
target, and obeys Jalilian-Marian-Iancu-McLerran-Weigert-Kovner
(JIMWLK) renormalization group
equations\cite{Jalilian-Marian-1997,Jalilian-Marian-PRD,Iancu-2001}. In a mean field approximation and large $N_c$ limit JIMWLK equation
is reduced to BK
equation~\cite{Balitsky1996,Balitsky1999,Kovchegov1999}.

The averaging over color sources can be done under MV model with a Gaussian weight functional. According to ref~\cite{cgc-NPA-2010},  the correlated two-gluon production can be expressed by uGD as

\begin{equation}\label{lead}
\left\langle\frac{dN_2^{\mathrm{corr.}}}{d^2\bo{p}_\perp dy_p d^2\bo{q}_\perp
dy_q}\right\rangle =\frac{
C_2}{\bo{p}^2_{\perp}\bo{q}^2_{\perp}}
    \int _0 ^{\infty}d^2 \bo{k}_\perp(D_1+D_2),
\end{equation}
where $C_2=\frac{\alpha^2_s N^2_c S_\perp}{4\pi^{10}(N^2_c-1)^3}$ and
\begin{eqnarray}\label{terms}
D_1= \Phi^2_{A}(y_p,\bo{k}_\perp)\Phi_{B}(y_p,\bo{p}_\perp-\bo{k}_\perp)D_B,\nonumber\\
D_2= \Phi^2_{B}(y_q,\bo{k}_\perp)\Phi_{A}(y_p,\bo{p}_\perp-\bo{k}_\perp)D_A,
\end{eqnarray}
with
\begin{equation}\label{eq:5}
D_{A(B)}=\Phi_{A(B)}(y_q,\bo{q}_\perp+\bo{k}_\perp)
+\Phi_{A(B)}(y_q,\bo{q}_\perp-\bo{k}_\perp).
\end{equation}
Here $\Phi_{A(B)}(y,\bo{k}_\perp)$ denotes uGD of projectile $A$ or target $B$. The single-gluon inclusive production reads
\begin{flalign}\label{eq:11}
&\left\langle\frac{dN_1}{d^2\bo{p}_\perp
dy_p}\right\rangle\nonumber\\
=&\frac{\alpha_s N_c
S_\perp}{\pi^4(N^2_c-1)}\frac{1}{\bo{p}^2_\perp}\int\frac{d^2\bo{k}_\perp}{(2\pi)^2}\Phi_{A}(y_p,\bo{k}_\perp)\Phi_{B}(y_p,\bo{p}_\perp
- \bo{k}_\perp).
\end{flalign}
The framework is valid to leading logarithmic accuracy in $x$ and momentum $p_\perp, q_{\perp}\gg Q_s$ and we only calculate the leading contributions in $p_\perp/Q_s$. 

The important ingredient in the above expressions is uGD ($\Phi$), which can be obtained by solving rcBK equation with a
given initial condition. To avoid repetition, details can be found
in ref~\cite{cgc-NPA-2010} and our previous
paper~\cite{zhao-1}. For p-p collision at 7 TeV, $Q^2_{s0}$ (with $Q_{s0}$ the initial value of $Q_s$ at $x_0$) is chosen to be 0.168 GeV$^2$~\cite{cgc-PRD-I}.

Based on the correlation function $C(\bo{p}_\perp,y_p;\bo{q}_\perp, y_q)$, the azimuthal correlation function is defined as
\begin{equation}
\begin{split}
C(\Delta\phi)=& \int^{p^{\mathrm{max}}_{\perp}}_{p^{\mathrm{min}}_{\perp}}  \frac{dp^2_{\perp}}{2} \int^{q^{\mathrm{max}}_{\perp}}_{q^{\mathrm{min}}_{\perp}} \frac{dq^2_{\perp}}{2} \\ & 
\times\int d\phi_p \int d\phi_q \delta(\phi_q-\phi_p-\Delta\phi)C(\bo{p}_\perp,y_p;\bo{q}_\perp, y_q),
\end{split}
\end{equation} 
which describes the correlation of two particles with rapidity $y_p$ and $y_q$ and their azimuthal separation $\Delta\phi$ in given transverse momentum intervals $(p^{\mathrm{min}}_{\perp}, p^{\mathrm{max}}_{\perp})$ and $(q^{\mathrm{min}}_{\perp}, q^{\mathrm{max}}_{\perp})$. We do not integrate over rapidity $y$ because we focus on correlations between two gluons in given rapidity locations. The rapidity gap is defined as $\Delta y=y_q-y_p$.

\section{Fine structure of azimuthal correlation and its transverse momentum and rapidity dependence}\label{sec:results}

\begin{figure*}[t]
\begin{centering}
\begin{tabular}{ccc}
$\vcenter{\hbox{\includegraphics[scale=0.45]{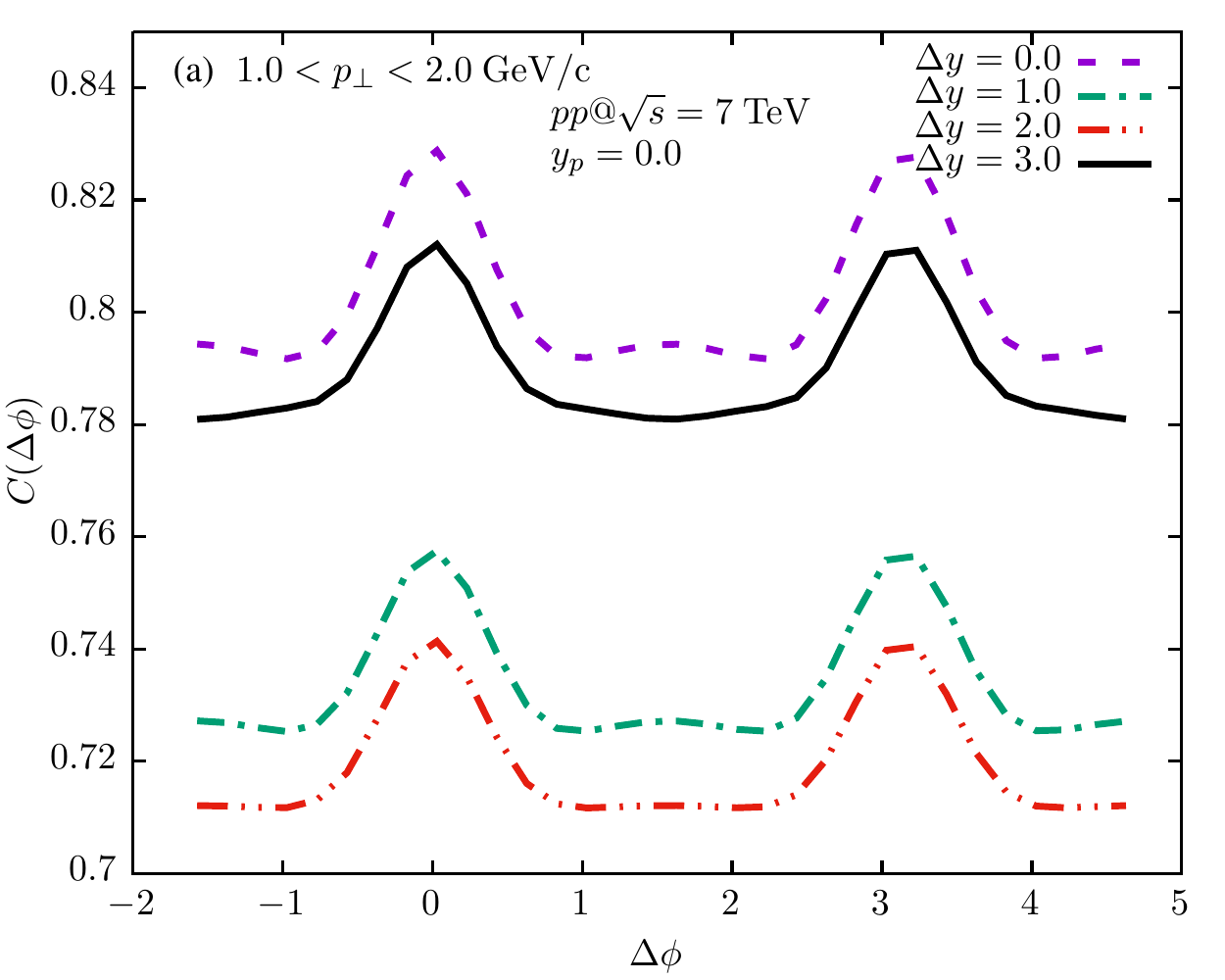}}}$ &
$\vcenter{\hbox{\includegraphics[scale=0.45]{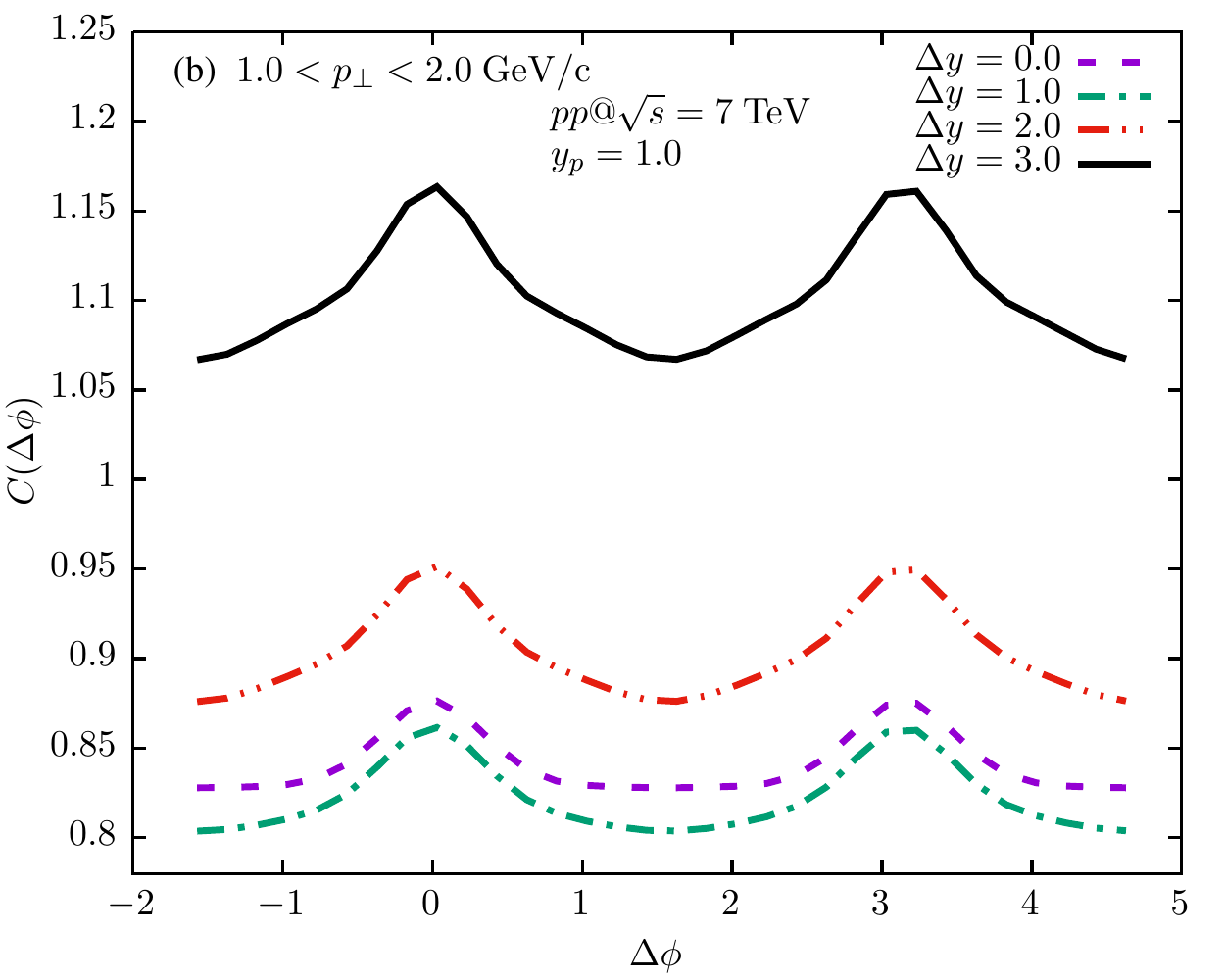}}}$
& $\vcenter{\hbox{\includegraphics[scale=0.45]{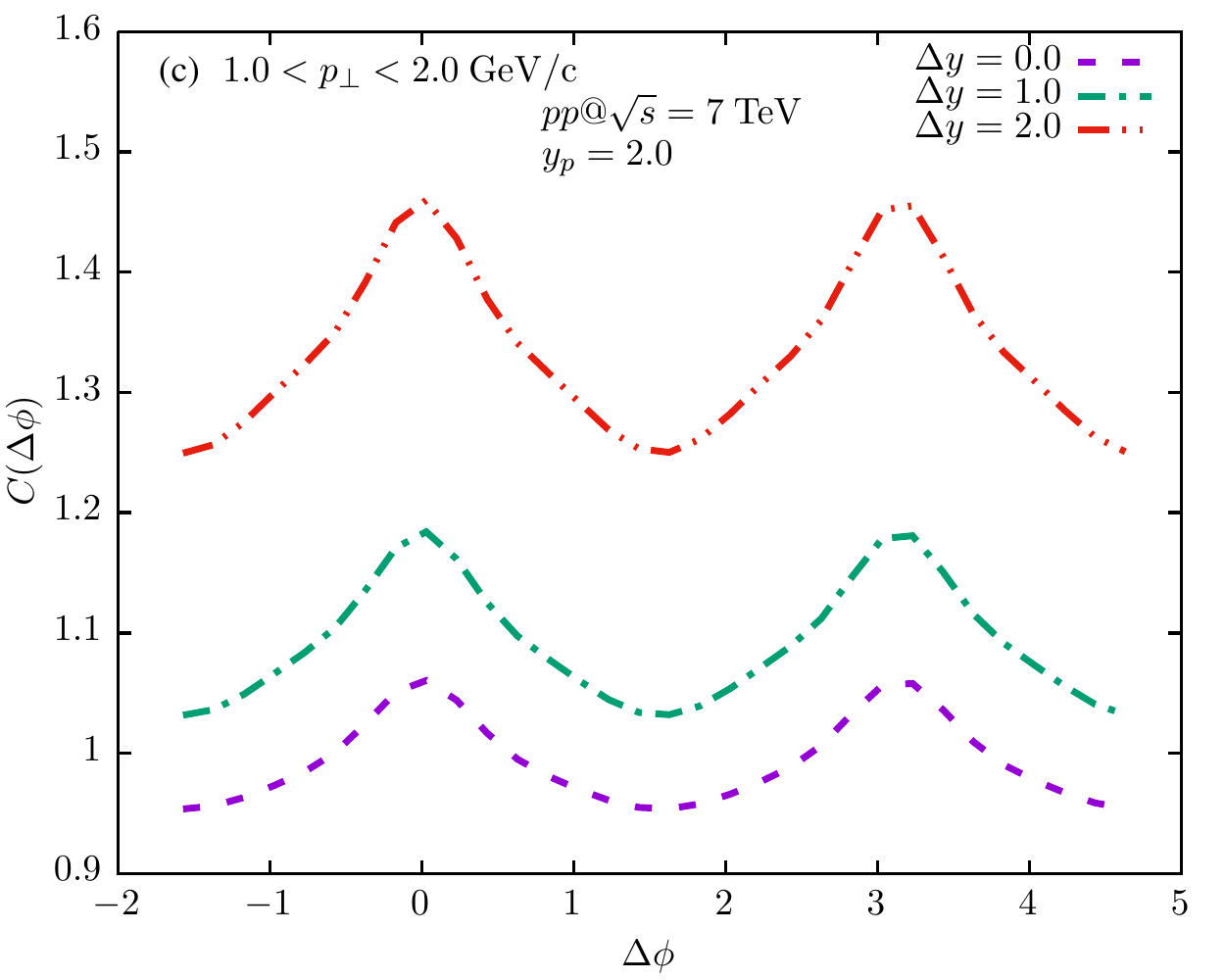}}}$ 
\end{tabular}
\par\end{centering}
\caption{Azimuthal correlation integrated in $1<p_{\perp}<2$GeV$/$c and $1<q_{\perp}<2$GeV$/$c is plotted as a function of $\Delta\phi$ for different rapidity gap $\Delta y=y_q-y_p$ when one gluon is located at (a) $y_p=0.0$, (b)$y_p=1.0$, (c)$y_p=2.0$, respectively.}
\label{pt-inte}
\end{figure*}

\begin{figure*}
\begin{centering}
\begin{tabular}{ccc}
$\vcenter{\hbox{\includegraphics[scale=0.45]{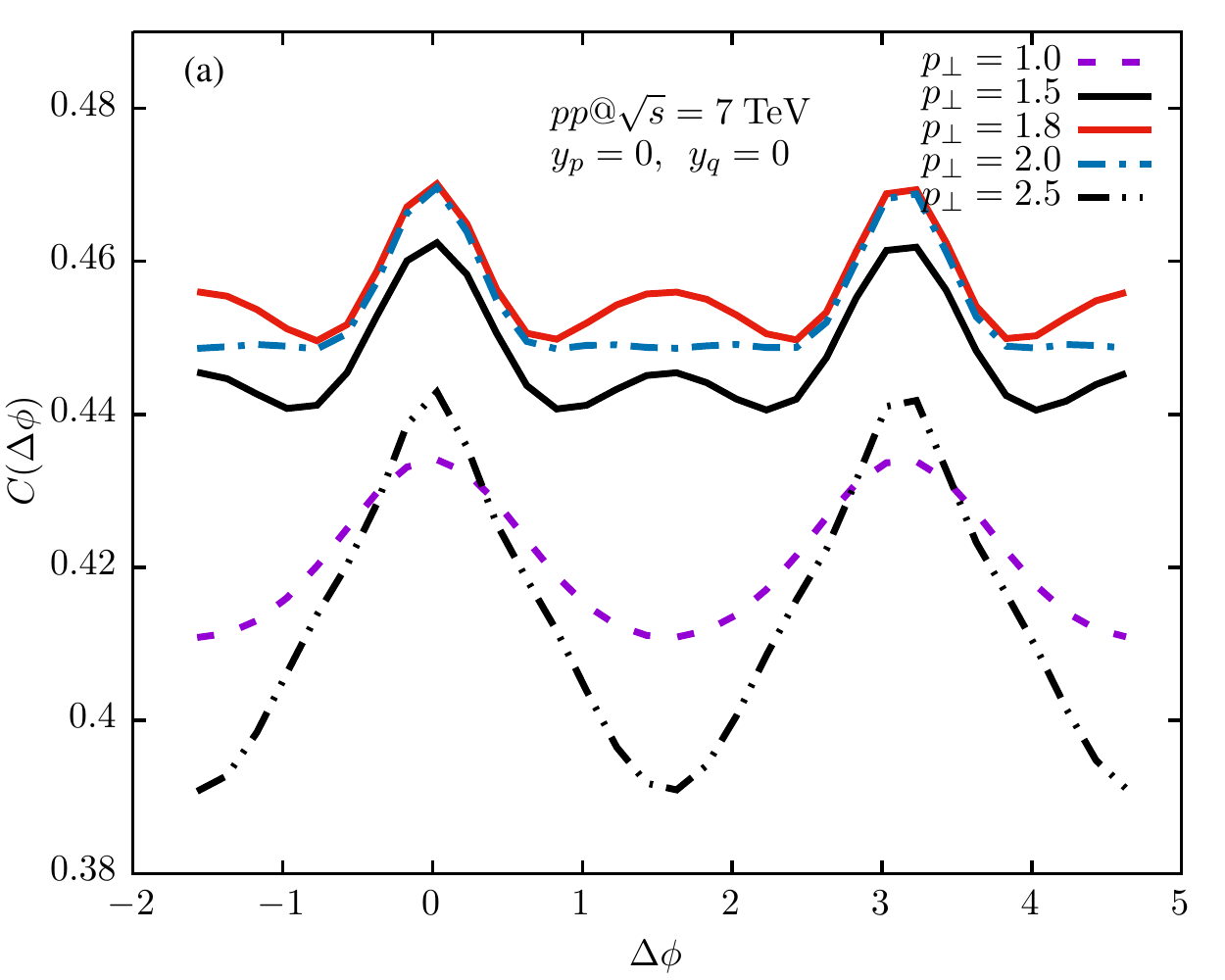}}}$ &
$\vcenter{\hbox{\includegraphics[scale=0.45]{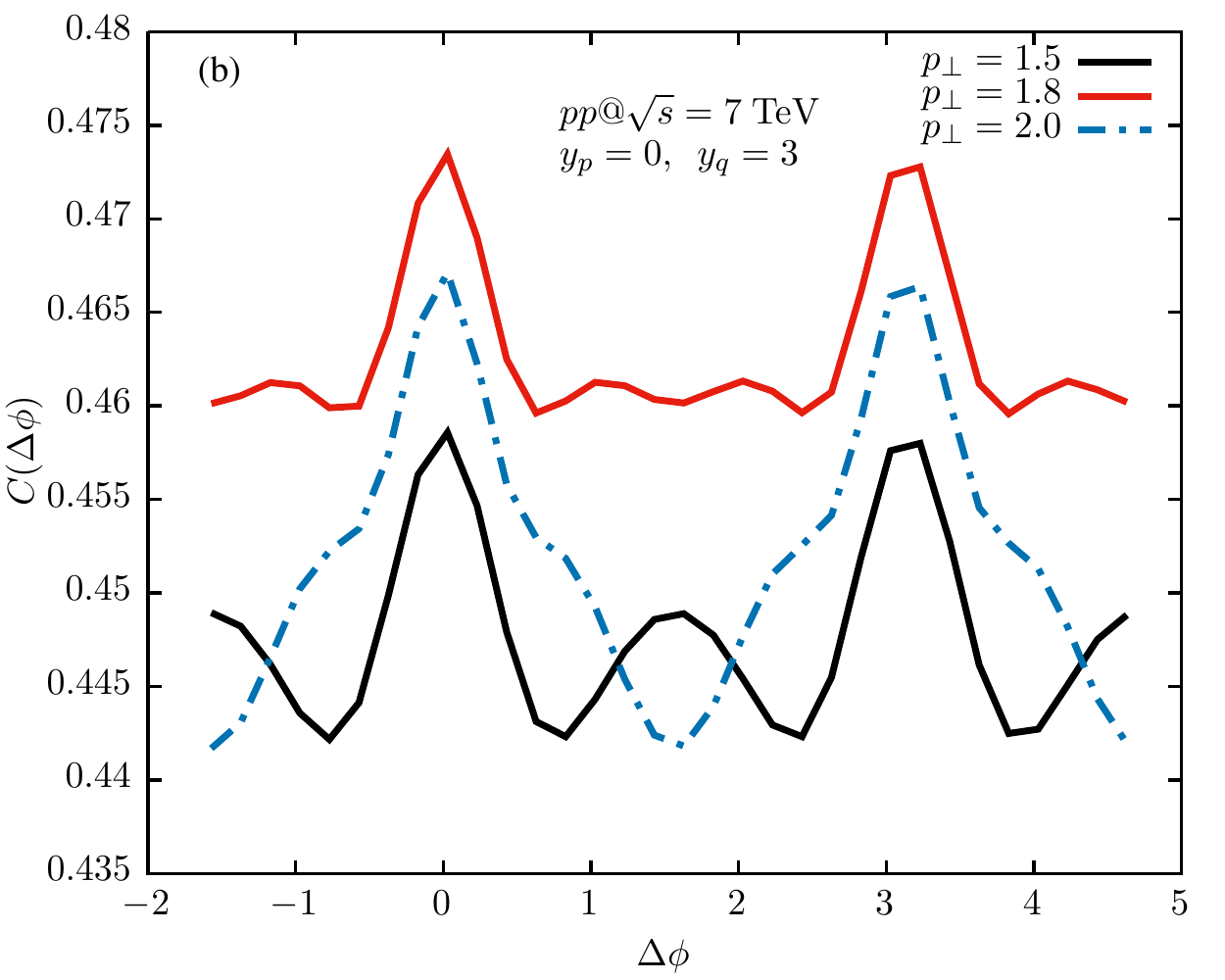}}}$
& $\vcenter{\hbox{\includegraphics[scale=0.45]{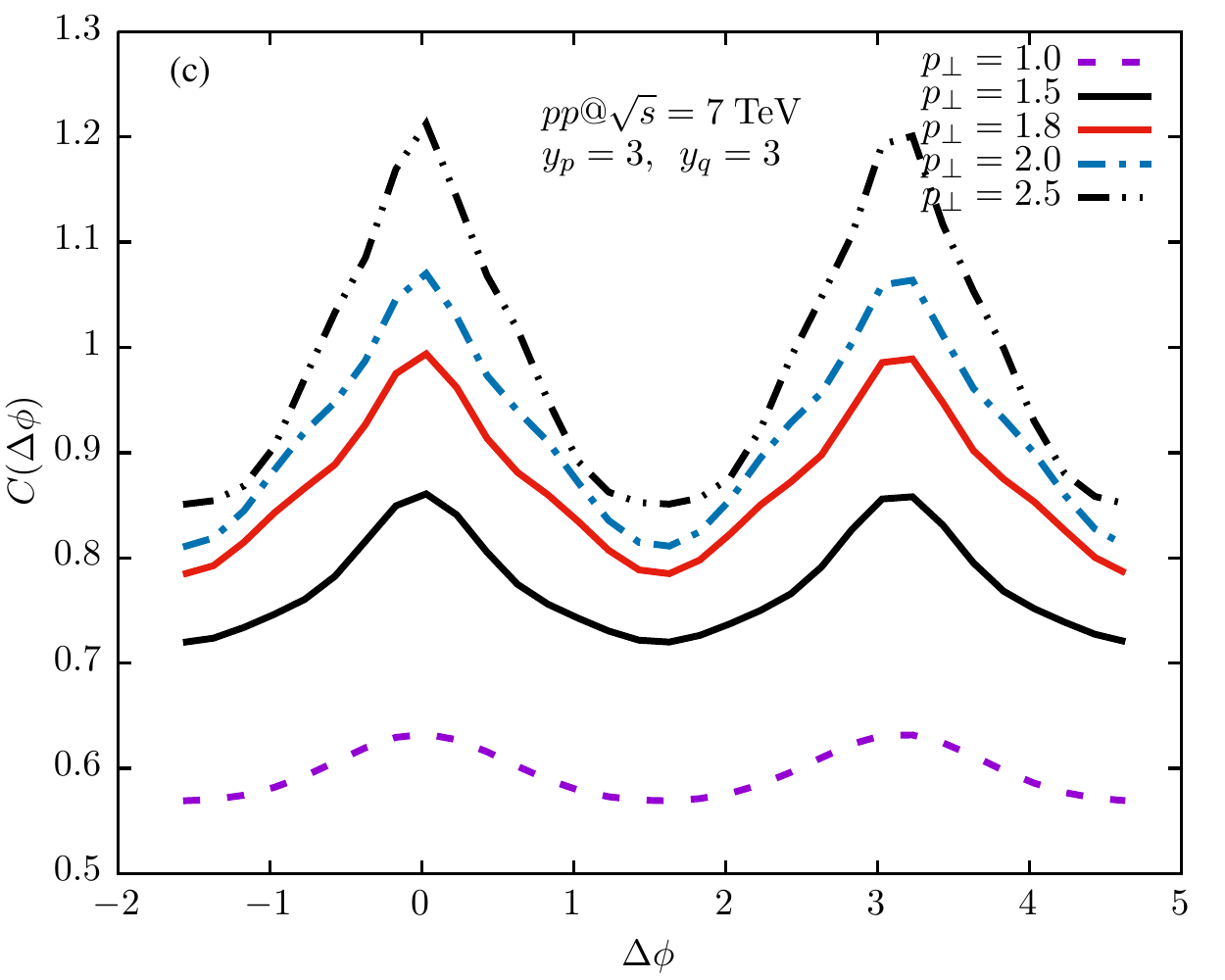}}}$ 
\end{tabular}
\par\end{centering}
\caption{Azimuthal correlation at five values of $p_{\perp}$ with $p_{\perp}=q_{\perp}$ at (a)$y_p=0, y_q=0$; (b)$y_p=0, y_q=3$; (c)$y_p=3, y_q=3$. In (b) the curves for $p_{\perp}=1.0$ and 2.5 GeV$/$c are similar to that in (a) and not shown in order to make the fine structure of other curves visible.}
\label{pt-dep}
\end{figure*}

Both experimental data~\cite{CMS-pp-JHEP-2010,ALICE-pPb-PLB-2016} and CGC~\cite{cgc-PRD-I,cgc-PRD-II,cgc-PRD-III}  show that the correlation at near side gets strongest in an intermediate $p_{\perp}$ interval, approximately $1<p_{\perp}<3$ GeV$/$c. A calculation from CGC points out that the correlation function gets maximum at $p_\perp\sim Q_{\mathrm{sA}}+Q_{\mathrm{sB}}=2Q_{\mathrm{sp}}=1.8$ GeV$/$c for minimum bias p-p collisions, where $Q_{\mathrm{sA(B)}}$ denotes the saturation momentum of projectile or target~\cite{zhao-1}. To obtain the strongest correlation, the correlation function is firstly calculated within $1<p_{\perp}<2$ GeV$/$c in this paper. 

To see how the azimuthal correlations change with rapidity gap and rapidity location, the azimuthal correlations are calculated at $\Delta y=0, 1, 2, 3$ with the rapidity location of one gluon chosen to be $y_p=0, 1, 2$, so that both small rapidity location and large rapidity location are included. The results are shown in \fig{pt-inte}. As mentioned in ref~\cite{cgc-PRD-III}, the azimuthal correlation from CGC shows a symmetric structure about $\pi/2$, i.e. one peak is located at near side $\Delta\phi=0$ and the other peak at away side $\Delta\phi=\pi$, which represent collimation production in CGC. Here we reproduce that structure in different rapidity locations, as all curves in \fig{pt-inte}(a), (b) and (c) show. Especially for long-range rapidity, e.g. $\Delta y=2.0$ as red curves show, correlations show two peaks at $\Delta\phi=0$ and $\pi$, which contributes to double-ridge phenomenon observed in data. 

When one gluon is located at small rapidity, e.g. $y_p=0.0$ shown in \fig{pt-inte}(a), the correlation strength at $\Delta\phi=0$ decreases when the rapidity gap increases from 0 to 2, and then rises significantly when the rapidity gap gets 3, which reproduces the trend of rapidity correlation at $\Delta\phi=0$ in ref~\cite{zhao-1} (Fig. 4 therein). The correlation strength at $\Delta y=0$ is rather high, compared to that of $\Delta y=1$ and 2, which is due to the contribution of short range correlation at $\Delta y=0$ from quantum evolution and is affected by the strength of running coupling~\cite{zhao-2}. The correlation strength at $\Delta y=3$ is rather high which is long range correlation in rapidity resulting from longitudinal boost invariance in the picture of color flux tubes of glasma~\cite{cgc-NPA-2008}. The dependence of correlation strength on the rapidity gap was discussed in detail in our previous papers~\cite{zhao-1, zhao-2} and not the focus in this paper. 

Besides the magnitude of correlations, the shape of the correlations as a function of $\Delta\phi$ shows interesting features. The two peaks at $\Delta\phi=0$ and $\pi$ exist in all cases in the following and the focus of this paper is the correlation structure between the two peaks. In \fig{pt-inte}(a), the curve around $\Delta\phi=\pi/2$ shows a moderate bump at $\Delta y=0$, which persists at $\Delta y=1$ and flattens at $\Delta y=2$, finally turns to be a shallow valley in the case of $\Delta y=3$. In \fig{pt-inte}(b), the curve around $\Delta\phi=\pi/2$ shows a flat structure at $\Delta y=0$ which reduces gradually to a valley at $\Delta y=1$, 2 and 3. In \fig{pt-inte}(c), all curves show a valley around $\Delta\phi=\pi/2$ no matter it is at short-range rapidity ($\Delta y=0$, 1) or long-range rapidity ($\Delta y=2$). Comparing the three sub-figures for different rapidity locations, the bump around $\Delta\phi=\pi/2$ is limited to small rapidity location.

Specifically, for azimuthal correlations at short-range rapidity $\Delta y=0$, as can be seen from the purple curves in \fig{pt-inte}(a), (b) and (c), a bump appears at  $\Delta\phi=\pi/2$ when $y_p=0$, and it disappears at $y_p=1$ and 2, which indicates that azimuthal correlations at short-range rapidity vary with the rapidity location of the chosen gluon. So do the azimuthal correlations at long-range rapidity, as can be seen from the red curves in \fig{pt-inte}(a), (b) and (c). It suggests that azimuthal correlations at the same rapidity gap change with the rapidity location of the chosen gluon.

In order to obtain the azimuthal correlations at different rapidity location in detail, we calculate the correlations between two gluons both at small rapidity, e.g. $y_p=0, y_q=0$ shown in \fig{pt-dep}(a), two gluons both at large rapidity, e.g. $y_p=3, y_q=3$ shown in \fig{pt-dep}(c), and one gluon at small rapidity and the other at large rapidity, e.g. $y_p=0, y_q=3$ shown in \fig{pt-dep}(b). Since azimuthal correlations are $p_{\perp}$ sensitive, an integration over $p_{\perp}$ may smear some correlation structure. Five values of $p_{\perp}$ are tried in the following calculations, i.e. $p_{\perp}=1.0, 1.5, 1.8, 2.0, 2.5$ GeV$/$c. For simplicity, $q_\perp$ is chosen to be equal to $p_\perp$ at first.

Correlation patterns are more diverse at single  $p_{\perp}$. In \fig{pt-dep}(a), the correlation at $p_{\perp}=1.0$ GeV$/$c shows a valley at $\Delta\phi=\pi/2$. As $p_{\perp}$ increases to 1.5 GeV$/$c a moderate bump at $\Delta\phi=\pi/2$ begins to appear and strengthens at $p_{\perp}=1.8$ GeV$/$c. As $p_{\perp}$ increases further to 2.0 GeV$/$c the correlation at $\Delta\phi=\pi/2$ drops to a flat structure and finally returns to a valley  at a larger $p_{\perp}$ of 2.5 GeV$/$c. In \fig{pt-dep}(b), the curves for $p_{\perp}=1.0$ and 2.5 GeV$/$c are similar to that in (a) and hence not shown in order to make the fine structures of other curves visible.  
The correlation patterns at $p_{\perp}=1.5$ GeV$/$c are similar to that in \fig{pt-dep}(a), while for $p_{\perp}=1.8$ GeV$/$c two bumps appear on the two sides of $\Delta\phi=\pi/2$, approximately at $\Delta\phi\approx1.0$ and $\Delta\phi\approx2.0$. Compared with the flat structure of $p_{\perp}=2.0$ GeV$/$c in \fig{pt-dep}(a), a valley appears in \fig{pt-dep}(b), with two shoulders on the two sides of $\Delta\phi=\pi/2$. It is worth noticing that the positions of the two shoulders are almost the same as that of the two bumps of $p_{\perp}=1.8$ GeV$/$c. However, all the bumps and flat structure existing in \fig{pt-dep}(a) and (b), which we call fine structures in the following, nearly disappear in \fig{pt-dep}(c), with only slight shoulders on the two sides of $\Delta\phi=\pi/2$ at  $p_{\perp}=1.8$ GeV$/$c and 2.0 GeV$/$c. 

The above mentioned phenomenon that azimuthal correlations at the same rapidity gap change with the rapidity location of the chosen gluon is more obvious in \fig{pt-dep}(a) and (c) where the rapidity gap is the same, i.e. $\Delta y=0$. By comparing the red curves in \fig{pt-dep}(a) and (c), we can see that the bump at $\Delta\phi=\pi/2$ only exist in small rapidity location. Furthermore, the bumps (one or two) around $\Delta\phi=\pi/2$ only exist in \fig{pt-dep}(a) and (b) which further indicates that these fine structures require at least one gluon located at small rapidity. Not only that, correlations calculated at single $p_{\perp}$ rather than integration in a wide $p_{\perp}$ range, help to obtain these patterns. Single $p_{\perp}$ at 1.5, 1.8 and 2.0GeV$/$c, i.e. a value near $p_\perp\sim 2Q_{\mathrm{sp}}=1.8$GeV$/$c, are most likely to show fine structures between $\Delta\phi=0$ and $\pi$. It means that azimuthal correlations have a sensitive range in transverse momentum, a rough interval between 1.5 and 2.0 GeV$/$c, which is associated with the saturation momentum of colliding particles.

In fact, the single bump at $\Delta\phi=\pi/2$ and the double bumps or shoulders at $\Delta\phi\approx1.0$ and $\Delta\phi\approx2.0$ represent two harmonic components in the azimuthal correlations. The single bump at $\Delta\phi=\pi/2$ represents a component of $\cos(4\Delta\phi)$ with its local maximum at $\Delta\phi=\pi/2$. The double bumps or shoulders represent a component of $\cos(6\Delta\phi)$ with its local maximum at $\Delta\phi=\pi/3\approx1.0$ and $\Delta\phi=2\pi/3\approx2.0$. The difference between double bumps and double shoulders lies in a large and small value of the coefficients of the sixth order harmonic component. In the same way, the main peaks at $\Delta\phi=0$ and $\Delta\phi=\pi$ represent a dominant component of $\cos(2\Delta\phi)$. If a Fourier expansion is applied to the azimuthal correlation function $C(\Delta\phi)$, it is natural to get the second order, fourth order and sixth order harmonic coefficients. High order harmonic components only get prominent when at least one gluon is located at small rapidity and has transverse momentum near 2 times the saturation momentum of colliding proton, i.e. $2Q_{\mathrm{sp}}=1.8$ GeV$/$c. 
 
The explanation of the patterns in azimuthal correlations depends on mechanism in this calculation. By the glasma graph shown in \fig{glasma-evolution}, the correlation function is proportional to the correlated two-gluon production and can be expressed by a convolution of four uGDs as shown in equations (\ref{lead}), (\ref{terms}) and (\ref{eq:5}). The integrand of equation (\ref{lead}) is explicitly written as 
\begin{eqnarray}\label{eq8}
D_1&= \Phi^2_{A}(y_p,\bo{k}_\perp)\Phi_{B}(y_p,\bo{p}_\perp-\bo{k}_\perp)\Phi_{B}(y_q,\bo{q}_\perp+\bo{k}_\perp)
\nonumber\\ &+\Phi^2_{A}(y_p,\bo{k}_\perp)\Phi_{B}(y_p,\bo{p}_\perp-\bo{k}_\perp)\Phi_{B}(y_q,\bo{q}_\perp-\bo{k}_\perp);\\
D_2&= \Phi^2_{B}(y_q,\bo{k}_\perp)\Phi_{A}(y_p,\bo{p}_\perp-\bo{k}_\perp)\Phi_{A}(y_q,\bo{q}_\perp+\bo{k}_\perp)\nonumber\\
&+\Phi^2_{B}(y_q,\bo{k}_\perp)\Phi_{A}(y_p,\bo{p}_\perp-\bo{k}_\perp)\Phi_{A}(y_q,\bo{q}_\perp-\bo{k}_\perp).
\end{eqnarray}
Since uGD ($\Phi$) peaks at $Q_s$, transverse momentum far from $Q_s$ contributes little to the correlation. In order to make a significant contribution to the correlation function, 
\begin{equation}
|\bo{k}_\perp|\sim Q_{\mathrm{s}},\quad |\bo{p}_\perp-\bo{k}_\perp|\sim Q_{\mathrm{s}}\quad \text{and}\quad |\bo{q}_\perp-\bo{k}_\perp|\sim Q_{\mathrm{s}}
\end{equation} 
are required simultaneously in the second term of both $D_1$ and $D_2$. If $\bo{p}_\perp$ and $\bo{q}_\perp$ are identical, and consequently $\bo{p}_\perp-\bo{k}_\perp=\bo{q}_\perp-\bo{k}_\perp$, the above conditions can be satisfied simultaneously which leads to collimation production at $\Delta\phi=0$. Similarly, significant contributions from the first term of $D_1$ and $D_2$ requires $\bo{p}_\perp$ and $\bo{q}_\perp$ are antiparallel which leads to collimation production at $\Delta\phi=\pi$. It explains why two gluons which are collimated have the most strongest correlations.

On the other hand, uGD only depends on the magnitude of transverse momentum, not on its direction. In that case, we should only require equal module of the two arguments, i.e. $|\bo{p}_\perp-\bo{k}_\perp|\sim|\bo{q}_\perp-\bo{k}_\perp|\sim Q_{\mathrm{s}}$. Two parallel or antiparallel transverse momenta are sufficient but unnecessary conditions for large correlations. According to the triangle law of vector addition, $\bo{k}_\perp$, $\bo{p}_\perp$ and $\bo{p}_\perp-\bo{k}_\perp$ constitute the three edges of a triangle. Since the maximum of uGD needs $|\bo{k}_\perp|\sim Q_{\mathrm{s}}$ and $|\bo{p}_\perp-\bo{k}_\perp|\sim Q_{\mathrm{s}}$, the angle between $\bo{k}_\perp$ and $\bo{p}_\perp$ (denoted as $\Delta\phi_{kp}$ which satisfies $\cos(\Delta\phi_{kp})=\frac{|\bo{k}_\perp|^2+|\bo{p}_\perp|^2-|\bo{p}_\perp-\bo{k}_\perp|^2}{2|\bo{k}_\perp| |\bo{p}_\perp|}$) is determined by $|\bo{p}_\perp|$. So does $\Delta\phi_{kq}$. Thus the relative azimuth $\Delta\phi=\phi_q-\phi_p$ at which correlations become significant depends on the magnitude of the two momenta. That explains why the peak position between $\Delta\phi=0$ and $\pi$ varies with transverse momenta, as shown in \fig{pt-dep}(a) and (b). Since $Q_\mathrm{s}$ is $y$ dependent, two edges in that triangle depdend on $y$. For given $p_\perp$, $\Delta\phi_{kp}$ varies with $y$ which means the relative azimuth at which correlations become significant depends on rapidity $y$.  At some momenta or rapidity, the correlations show a simple structure of a valley between $\Delta\phi=0$ and $\pi$, without bumps and shoulders. That is because the length of the three edges are not appropriate to form a triangle and hence the uGDs can not get maximum simultaneously to develop a large correlation.

\begin{figure}
\includegraphics[scale=0.5]{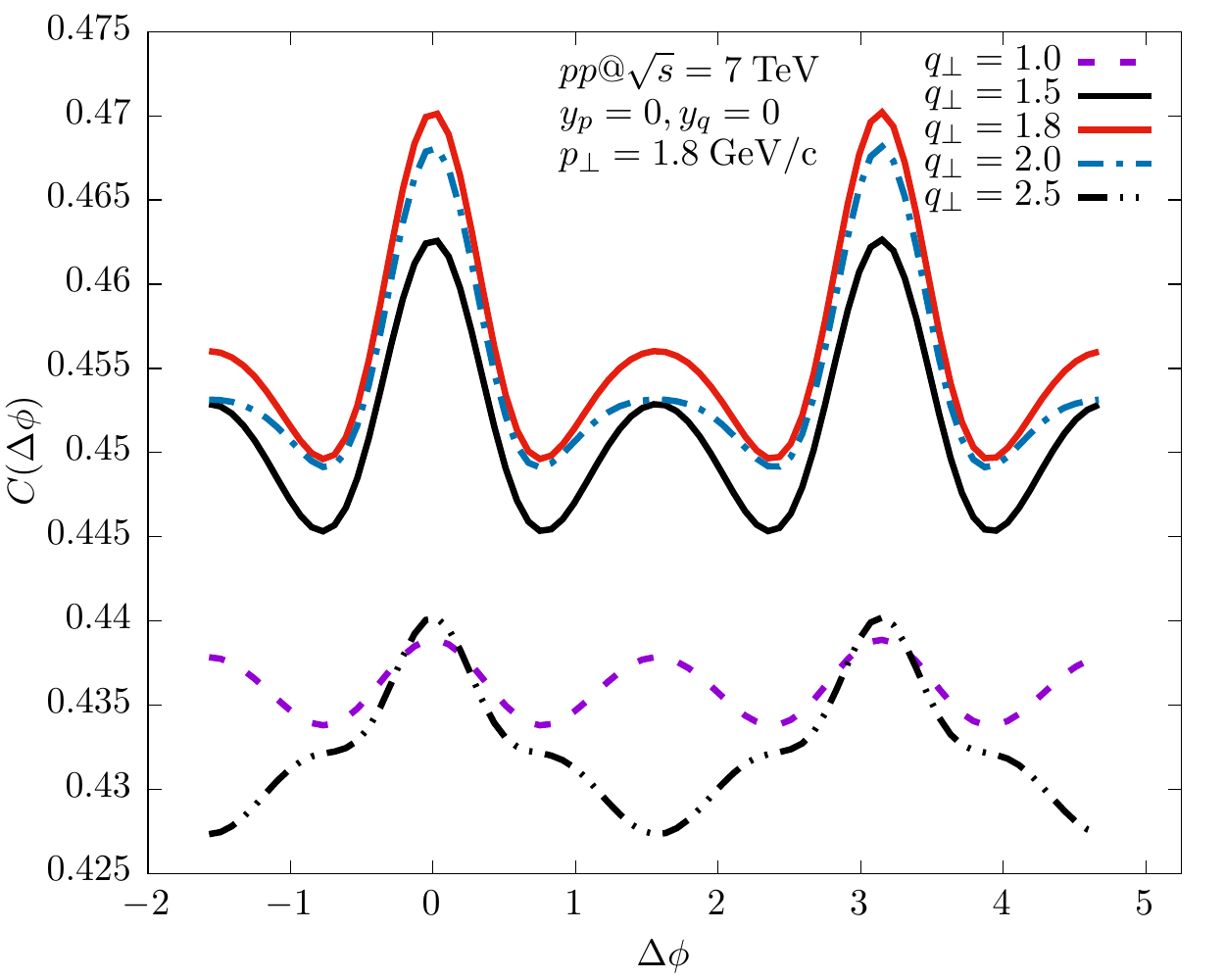}
\caption{Azimuthal correlations at five values of $q_{\perp}$ with $p_{\perp}=1.8$GeV$/$c at $y_p=0, y_q=0$.}
\label{pt-qt}
\end{figure}

From this kind of view, we can infer that if $p_\perp$ is fixed, the azimuthal correlation pattern for different $q_\perp$ should be different. This is indeed the case, as shown in \fig{pt-qt}. To observe the fine correlation patterns, $p_\perp$ is chosen as 1.8 GeV$/$c, and $y_p=0, y_q=0$ with both gluons at small rapidity. When $q_\perp$ varies from 1.0 to 2.0 GeV$/$c, the peak at $\Delta\phi=\pi/2$ always exists despite of differences of its strength. When $q_\perp$ departs from the sensitive range, the peak reduces to double shoulders. It indicates that these correlation patterns require that at least one gluon has transverse momentum within sensitive range.
   
\section{Summary and outlook}\label{sec:summary}

In this paper we study two-gluon azimuthal correlations and their $p_\perp$ and $y$ dependence by using the CGC formalism. We find that two-gluon azimuthal correlations are sensitive to the detailed dynamical features of CGC. Here, two gluons are chosen from small rapidity location, or large rapidity location, or one is from small rapidity location and the other is from large rapidity location. Results show that azimuthal correlations at the same rapidity gap change with the rapidity location of the chosen gluon. Fine structures around $\Delta\phi=\pi/2$, i.e. bumps or shoulders, show up when at least one gluon is located at small rapidity, which suggests that fine structures of correlation patterns are specific to small $x$ region. Single value of $p_\perp$ near $Q_{\mathrm{sA}}+Q_{\mathrm{sB}}=2Q_{\mathrm{sp}}=1.8$ GeV$/$c, instead of integration over $p_\perp$ in a wide range like $1<p_\perp<2$GeV$/$c,  are more likely to illustrate fine structures between  $\Delta\phi=0$ and $\pi$. The fine structures around $\Delta\phi=\pi/2$ correspond to high order harmonic components in a Fourier expansion. That means high order harmonic components only get prominent when at least one gluon is located at small rapidity and at least one gluon has transverse momentum near $2Q_{\mathrm{sp}}$ (the sum of the saturation momentum of two colliding particles). 

The reason why rapidity location has an influence on azimuthal correlations may be that different rapidity locations correspond to different $x$. The fact that fine structures of azimuthal correlations only exist at small rapidity may indicate that small $x$ evolution should be responsible for the fine structures of azimuthal correlations and these high order harmonic components observed here.

Furthermore, single value of $p_\perp$ is impractical for real analysis. Instead, a narrow bin of $p_\perp$, much less than 1 GeV$/$c, will help to see the fine structure, otherwise the bumps will be less noticeable. Even so, one must be careful to examine the fine structures because the magnitude of fine structures is much less than the amplitude of the main peaks at $\Delta\phi=0$ and $\pi$.

The analysis method here, i.e. finer binning of transverse momentum near saturation scale and focusing on small rapidity location, provides a way to address the properties of the initial glasma state. Assuming that final state interactions do not change the pattern dramatically, it is interesting to see if the dependence of azimuthal correlations on rapidity location and transverse momentum exist in experimental data. It may serve as a proof of glasma dynamics.

\section*{Acknowledgement}
This work is supported in part by the Major State Basic Research Development Program of China under Grant No. 2014CB845402, the Ministry of Science and Technology (MoST) under grant No. 2016YFE0104800, the NSFC of China under Grant No. U1732271 and 11647093. 

\providecommand{\href}[2]{#2}\begingroup\raggedright\endgroup

\end{document}